\documentclass[11pt, letter]{article}
\usepackage[utf8]{inputenc}
\usepackage[letterpaper, margin=1in]{geometry}

\usepackage[final]{microtype}

\usepackage{booktabs}
\usepackage{graphicx}
\usepackage[hyphens]{url}
\usepackage[hidelinks]{hyperref}

\usepackage{xcolor}
\usepackage{listings}

\definecolor{mGreen}{rgb}{0,0.6,0}
\definecolor{mGray}{rgb}{0.5,0.5,0.5}
\definecolor{mPurple}{rgb}{0.58,0,0.82}
\definecolor{backgroundColour}{rgb}{0.95,0.95,0.92}

\newcommand{\code}[1]{\texttt{#1}}

\usepackage[inline]{enumitem}

\usepackage{authblk}

\usepackage{comment}

\usepackage[numbers]{natbib}
\begin{document}
\renewcommand{\labelenumi}{(\arabic{enumi})}
\renewcommand{\labelenumii}{(\alph{enumii})}

\title{Making existing software quantum safe: a case study on IBM Db2}

\author[1]{Lei Zhang}
\author[1]{Andriy Miranskyy}
\author[2]{Walid Rjaibi} 
\author[2]{Greg Stager} 
\author[3]{Michael Gray} 
\author[4]{John Peck} 

\affil[1]{Department of Computer Science, Toronto Metropolitan University, Toronto, Canada}
\affil[2]{IBM Canada Lab, Markham, Canada}
\affil[3]{IBM, Bundall, Australia}
\affil[4]{IBM, Austin, USA}
\affil[ ]{{\{leizhang, avm\}@torontomu.ca}, \{wrjaibi, gstager\}@ca.ibm.com, mickgray@au1.ibm.com, johnpeck@us.ibm.com}

\date{}

\maketitle

\begin{abstract}
The software engineering community is facing challenges from quantum computers (QCs). In the era of quantum computing, Shor's algorithm running on QCs can break asymmetric encryption algorithms that classical computers practically cannot. Though the exact date when QCs will become ``dangerous'' for practical problems is unknown, the consensus is that this future is near. Thus, the software engineering community needs to start making software ready for quantum attacks and ensure quantum safety proactively.

We argue that the problem of evolving existing software to quantum-safe software is very similar to the Y2K bug. Thus, we leverage some best practices from the Y2K bug and propose our roadmap, called 7E, which gives developers a structured way to prepare for quantum attacks. It is intended to help developers start planning for the creation of new software and the evolution of cryptography in existing software.

In this paper, we use a case study to validate the viability of 7E. Our software under study is the IBM Db2 database system. We upgrade the current cryptographic schemes to post-quantum cryptographic ones (using Kyber and Dilithium schemes) and report our findings and lessons learned. 

We show that the 7E roadmap effectively plans the evolution of existing software security features towards quantum safety, but it does require minor revisions. We incorporate our experience with IBM Db2 into the revised 7E roadmap.

The U.S. Department of Commerce's National Institute of Standards and Technology is finalizing the post-quantum cryptographic standard. The software engineering community needs to start getting prepared for the quantum advantage era. We hope that our experiential study with IBM Db2 and the 7E roadmap will help the community prepare existing software for quantum attacks in a structured manner.

\end{abstract}

\section{Introduction}\label{sec:intro}

The field of quantum computing is still young, but it has been evolving rapidly during the last decade. With the vast increase in computing power, quantum computers (QCs) promise to revolutionize many fields, including artificial intelligence, medicine, and space exploration~\citep{zhang2021quantum}. However, they can also be abused to break most common encryption algorithms, which modern cryptography depends upon today to ensure digital information's safety and privacy. Thus, we position that the software engineering community should start thinking about the impact of quantum computing on cybersecurity and the best practices to address these concerns. First, let us look at the evolution of QCs.

\subsection{The timeline of QCs}

The field of quantum computing is young. In 1982, Richard Feynman introduced the idea of simulating quantum physics with computers~\citep{feynman1982simulating}. It took a while to implement an actual QC. A partnership between academia and IBM created the first working QC with two qubits\footnote{Qubit is the basic unit of quantum information, on which QC operate. At any point of execution, the state of a classical computer (CC) is given by a vector of bits taking the values of $0$ and $1$. The state of a QC is, however, given by a vector of qubits and bits.} in 1998~\citep{Chuang1998}. Then, it took the company 18 years to develop a 5-qubit QC, which became accessible to the public in 2016~\citep{ibm2016}. 

At present, a few QCs are commercially available. D-Wave started selling adiabatic QC\footnote{The debate about adiabatic QC being a ``true'' QC is ongoing~\citep{albash2017}. A hybrid of adiabatic and gate-based QC is promising~\citep{barends_digitized_2016}, but no commercial implementation is available.} in 2011, with the current offerings having $>$~5000 qubits~\citep{DWaveAnn7:online}. Recently, D-Wave announced that it would follow the lead of IBM, Google, and others to develop an all-purpose quantum computer using a ``gate-model'', which has a similar design to those from its rivals~\citep{Onceapio1:online}. 

QCs are also available via fully-managed Cloud services. IBM gave access to  20- and 50-qubit gate-based superconducting QCs to academic and industrial partners to explore practical applications in 2017~\citep{ibm2017}, and a 65-qubit machine was offered in 2020~\citep{IBMsRoad67:online}. For non-commercial use, IBM offers 5- and 15-qubit QCs via IBM Q Experience online platform based on IBM Cloud (along with local- and Cloud-based simulators)~\citep{ibm_quantum}. Rigetti offered 8-qubit superconducting QC in 2017~\citep{RigettiC96:online}. Google built a 72-qubit gate-based superconducting QC in 2018~\citep{google2018}. IonQ introduced ion-trapped 11-qubit QC in 2019~\citep{Wright2019}. Honeywell created ion-trapped 10-qubit QC in 2020~\citep{QuantumC77:online}. Xanadu offered 8- and 12-qubit photonic QCs in 2020~\citep{XanaduRe21:online,CloudPla55:online}. Microsoft provides access to a simulator\footnote{A QC can be simulated on a CC~\citep{ibm_quantum,ms_quantum}. A quantum simulator interprets a mathematical function as part of a physical model~\citep{Johnson2014}; however, it will not yield performance improvement that a QC would provide, as the underlying host system of the simulator is still based on bits rather than qubits (a basic unit of quantum information). Thus, one needs a real QC to reap performance benefits.} of a topological QC via Microsoft Quantum Development Kit~\citep{ms_quantum} (and is planning to give access to an actual QC in the future).

Despite the rapid development of QCs, we conjecture that QCs will not replace CCs in the short run. Instead, QCs will be integrated into a System of Systems via private or public Cloud service~\citep{Miranskyy019,Miranskyy0D20,miranskyy2021testing}. Aggregated Cloud services are starting to appear as well. For example, Amazon Web Services started offering access to QCs from various vendors via its Braket service in 2019~\citep{aws_braket_intro}. Currently, it offers D-Wave adiabatic 2048- and 5640-qubit QCs, IonQ trapped-ion-based 11-qubit QC, and Rigetti 32-qubit superconducting QC~\citep{AmazonBr86:online}. 

To summarize, many formidable competitors are scaling up various QC architectures, bringing us closer to the day when QCs can solve practical problems. So let us look at when it may happen.

\subsection{Quantum advantage}

In the future, a large QC can solve some problems that a CC practically cannot, which is called quantum advantage~\citep{feynman1982simulating} (often used interchangeably with the term of quantum supremacy). Quantum advantage was demonstrated on a superconducting QC in 2019~\citep{arute2019quantum}. Another demonstration of quantum advantage was done on a photonic QC in 2020~\citep{zhong2020quantum}, although the setup used in the experiment may be difficult to scale up or generalize~\citep{choi2021}.

But when will QCs start solving real-world problems? Using existing QCs, quantum chemists are already able to improve simulations of small chemical systems~\citep{kandala2017hardware} and some large ones, albeit with approximations~\citep{robert2021resource}. However, quantitative financists will need a machine with $\approx$~7.5K logical qubits to price financial instruments~\citep{chakrabarti2020threshold}. Hackers will need a computer with 20M qubits to break the 2048-bit RSA key in less than a day~\citep{gidney2019factor}. 

To start addressing practical use-cases within the next decade, IBM stated that they ``need to at least double the Quantum Volume (QV) of our quantum computing systems every year.''~\citep{chow2020} So far, IBM is on track, demonstrating the QV of 64 on a QC with 27 qubits in 2020~\citep{IBMsRoad67:online}. By 2023, IBM plans to ship a QC of 1,121 qubits (with the expectation of proportional QV growth)~\citep{IBMsRoad67:online}.

\subsection{Quantum advantage: impact on cybersecurity}
The power of quantum computing will threaten modern cybersecurity platforms by speeding up 
\begin{enumerate*}
    \item  factorization of integers, solving the discrete logarithm problem, and the elliptic-curve discrete logarithm problem (using Shor’s algorithm~\citep{shor1997}); as well as 
    \item the search in a set  (with the help of Grover’s algorithm~\citep{grover1996fast}). 
\end{enumerate*}
Both tasks are foundational for modern encryption algorithms.

\begin{figure}
	\centering
	\includegraphics[width=\columnwidth]{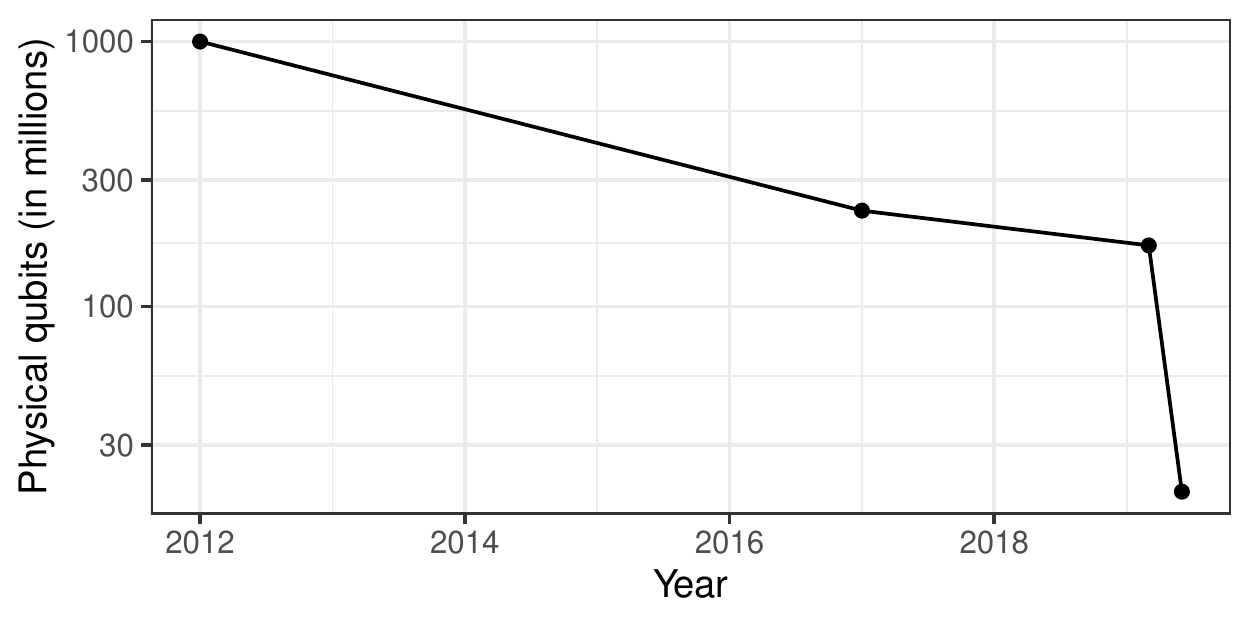}
	\caption{Number of qubits (in millions) needed to break RSA 2048 based on the literature review in~\cite[Table II]{gidney2019factor}. The x-axis denotes a year when an article was published.
	}
	\label{fig:20milli}
\end{figure}

While the existing QCs are not ready to use Shor’s algorithm for production problems, many brilliant people (both on the hardware and the algorithm sides) are racing to get us there. For example, Gidney et al.~\citep{gidney2019factor} combine a set of clever tricks and techniques to implement Shor’s algorithm using modern QC architecture. They show that one needs 20M qubits to break the 2048-bit RSA key in less than a day. They compare their solution against an existing one by Fowler et al.~\citep{Fowler2012Surface}, which was introduced in 2012. This baseline approach required one billion qubits. Thus, in seven years, algorithms designers reached a remarkable 165x~\cite[Table II]{gidney2019factor} improvement in the algorithm implementation. This dynamic is represented graphically in Fig.~\ref{fig:20milli}. The 20M qubits is not an optimal boundary, which implies that in the future, improvement schemes may be derived, further reducing the hardware requirements.

To improve QCs, electrical engineers and physicists will continue working on enhancing coherence, noise reduction, and error correction solutions. Theoretically~\citep{qubits4099}, a QC with 4099 ideal, i.e., perfectly stable, qubits can break RSA 2048 in 10 seconds. Such stable qubits will appear in  fault-tolerant quantum computers (FTQCs), which will employ error-correcting schemes that will allow us to operate on logical/ideal qubits (constructed from groups of physical qubits)~\citep{chao2018fault}. Small FTQCs may appear within ten years~\citep{sevilla2020forecasting}. Moreover, new contenders in this field (such as photonic QC developed by Xanadu~\citep{xanadu} or ionic QCs developed by IonQ~\citep{Wright2019} and Honeywell~\citep{QuantumC77:online}) may introduce new and efficient computing architectures that will speedup arrival of FTQCs. IonQ, for example, already produced a QC with 20 ideal qubits~\cite{lubinski2021application}. This paper will use the term FTQC to refer to a QC that is powerful enough to break existing encryption schemes.

\subsection{Quantum advantage: call to action}\label{sec:cta}

Since FTQCs are years away, can we still rely on existing encryption schemes without upgrading them? Unfortunately, the short answer is ``No.'' To arrive at this conclusion, let us examine the relationships between three time intervals:
\begin{enumerate}
    \item Time (denoted by $X$) needed to deploy quantum-safe cryptographic solutions;
    \item Time (denoted by $Y$) required to maintain the security of your encrypted data; and
    \item Time (denoted by $Z$) when a quantum computer capable of breaking existing encryption schemes, such as FTQC, will appear.
\end{enumerate}

Suppose  $X > Z$. In this scenario, the the deployment of quantum-safe cryptography (i.e., $X$) will take longer than the emergence of FTQCs (i.e., $Z$). QCs capable of breaking public-key encryption schemes may be developed within the next twenty years~\citep{PostQuan1:online}. While this feels like a long time, historically, it takes almost two decades to deploy public-key cryptography infrastructure~\citep{PostQuan1:online}. Thus, even if we begin planning to switch from existing encryption algorithms to quantum-safe ones today, upgrading the security features will take a lot of time and effort.

What if $X \leq Z$? It implies that we have updated our encryption schemes proactively, which is an essential topic of this paper.

Now, suppose $Y > Z$. In this scenario, we should maintain the security of the encrypted data (i.e., $Y$) past the arrival of FTQCs (i.e., $Z$). Malicious entities can, for example, harvest sensitive data that must remain confidential for many decades (e.g., state secrets) and then use that data to perform a harvest-then-decrypt attack. Therefore, an adversary can harvest sensitive data communications today and~--- when FTQCs are available~--- use that computing power to break today's non-quantum-safe encryption to access such sensitive data.

In the case of $Y \leq Z$, the data becomes stale and non-sensitive before the FTQCs arrive. For the time being, most data types fall into this category. However, new data are being generated every day, while $Z \to 0$. Thus, more data types will be at risk as time passes.

In summary, due to the uncertainties inherent in quantum threats and the ambiguous timeline, the sooner we start working on upgrading to quantum-safe encryption schemes, the sooner we will be able to mitigate this risk.

Now is the time to start planning the evolution of encryption schemes (e.g., identification of vulnerable hardware and software components and design of replacement schemes). It may take many months (if not years) to identify all problematic hardware and software components in an enterprise environment and devise corrective actions. An organization that begins working on these tasks now may be able to complete them by the time the security standards are finalized.

\subsection{Quantum-safe: existing solutions}\label{sec:existing-solutions}

Existing approaches proposed to ensure information security in the era of quantum computing have two trends. The first approach is quantum cryptography, which harnesses quantum mechanics to protect confidential internet communications. For example, quantum key distribution (QKD) is the most popular approach, where data is transferred via no-change and no-cloning photons instead of bits~\citep{RevModPhys.81.1301}. However, the downside of quantum cryptography is that it needs quantum channels for sending quantum bits among different locations~\citep{mosca2018cybersecurity}. The second approach is post-quantum cryptography (PQC)~--- the focus of this paper~--- which is about preparing for the era of quantum computing by updating existing mathematical algorithms and encryption standards with classical cryptographic algorithms that are secure against both QCs and CCs~\citep{Bernstein017,Song14,BuchmannBGP16}.

The standard of PQC is still in development. Based on the timeline of the National Institute of Standards and Technology (NIST), we expect to see a draft of the standard between 2022 and 2024~\citep{nist_timeline}. Nevertheless, it is still valuable to plan to upgrade existing cryptographic systems, particularly doing so while adhering to crypto-agility practices (as recommended by NIST~\citep{chen2016report}), so that any algorithm chosen today can be changed in the future (if necessary) without incurring a considerable cost. While we need to wait for one to three years to see the standard, we already have a good understanding of the classes of algorithms that will appear in this standard. In July 2020, NIST selected 15 candidates algorithms (7 finalists and 8 alternates) in Round 3 of the competition~\citep{nist_timeline}. 

This paper extends our previous publication~\citep{zhang2021quantum}, which introduced the 7E roadmap for software developers to address encryption-related challenges associated with quantum advantage. The \textbf{contributions} of this paper are 
\begin{enumerate}
    \item Validating the 7E roadmap by conducting a case study on IBM software, 
    \item Describing our PQC implementation for the IBM Db2 software\footnote{Our PQC schemes are generic and applicable to other software products.},
    \item Summarizing challenges and lessons learned based on our experience, and 
    \item Logging our best practices and improving the 7E roadmap.
\end{enumerate}
These contributions may be helpful to practitioners planning PQC upgrades and academics working in the area of software maintenance and evolution.

The rest of this paper is structured as follows. Section~\ref{sec:challenges} analyzes the impact of quantum computing on existing software systems. Section~\ref{sec:method} recaps the original 7E roadmap. We propose the enhanced 7E roadmap in Section~\ref{sec:new7e}. The enhancements are based on the case study, which is described in the following sections. Section~\ref{sec:industry} presents the industrial settings of our case study. Section~\ref{sec:experiment} illustrates the implementation of PQC and the performance evaluation. Section~\ref{sec:lessons} discusses the challenges we met and our takeaway messages for other researchers and practitioners. We discuss threats to validity and our concerns in Section~\ref{sec:discussions}. Finally, Section~\ref{sec:conclusion} concludes the paper. 

\section{The impact of quantum computing on existing systems}\label{sec:challenges}
It is essential to understand that quantum computing will affect encryption schemes differently depending on the class of encryption algorithm. In this section, we first elaborate on the asymmetric (public-key) encryption algorithms, which are used in many areas ranging from the Transport Layer Security (TLS) protocol (used to safeguard data passed between two systems on the Internet) to Pretty Good Privacy software used to encrypt and decrypt a file and safely transfer it between computers.  

We then proceed to the symmetric (private-key) encryption algorithms, such as Advanced Encryption Standard (AES), used to protect sensitive data. There exist numerous use-cases, ranging from encrypting a file archive (e.g., implemented in 7z and WinZip software) to encrypting computers' disks (e.g., using Apple MacOS FileVault and Symantec Endpoint Encryption).

\subsection{Asymmetric encryption}\label{sec:asymmetric}

\begin{table*}[t]
\centering
  \caption{Effective security strength of key encryption algorithms as per~\citep{mavroeidis2018impact}}
  \label{tab:key}
  \resizebox{\textwidth}{!}{
  \begin{tabular}{lrrr}
    \toprule
    \multicolumn{1}{c}{Encryption algorithm} & \multicolumn{1}{c}{Key size (bits)} & \multicolumn{1}{c}{Effective security level on CCs (bits)} & \multicolumn{1}{c}{Effective security level on QCs (bits)} \\
    \midrule
    RSA 1024 & 1024 & 80 & 0 \\
    RSA 2048 & 2048 & 112 & 0 \\
    ECC 256 & 256 & 128 & 0 \\
    ECC 384 & 384 & 256 & 0 \\
    AES 128 & 128 & 128 & 64 \\
    AES 256 & 256 & 256 & 128 \\
  \bottomrule
\end{tabular}
}
\end{table*}

Asymmetric encryption algorithms, which are based on factoring large integers (e.g., Rivest-Shamir-Adleman~--- RSA), discrete logarithms (e.g., Elliptic Curve Cryptography~--- ECC, and Diffie-Hellman key exchange~--- DH), or similar approaches (see~\citep{mavroeidis2018impact} for review)  will need to be replaced by quantum-safe alternatives~\citep{mavroeidis2018impact}. Effective security strength, shown in Table~\ref{tab:key}, suggests that the strength of the RSA and ECC is somewhat weaker or comparable to AES on a CC, but is extremely weak on a QC. This is because Shor’s algorithm can perform integer factorization in polynomial time; so what requires thousands of years with classical computers would only take days/hours on a large-scale quantum computer. This, of course, assumes that a large-scale quantum computer with the required number of qubits and degree of coherence exists, which is not the case right now. 

\subsection{Symmetric encryption}\label{sec:symmetric}
Unlike asymmetric encryption algorithms, symmetric encryption algorithms do not face an existential threat: one needs to perform a brute-force attack to break it. However, on a CC generation of $n$ keys require $O(n)$ operations, while on a QC it can be done using $O\left(\sqrt{n}\right)$ operations, thanks to Grover’s algorithm. Thus, a large quantum computer running Grover’s algorithm could provide a quadratic improvement in brute-force attacks on symmetric encryption algorithms, such as AES. This translates into a need to double key size to support the same level of protection. For AES specifically, this means using 256-bit keys to maintain today’s 128-bit security strength,\footnote{That is, an $n$-bit AES cipher provides a security level of $n/2$  because $\sqrt{2^n} = 2^{n/2}$.} as depicted in Table~\ref{tab:key}.

\subsection{Challenges on existing systems}\label{sec:impact}

If the legacy system is well-designed and actively maintained, then the solution is straightforward: one can replace an existing asymmetric algorithm with a new one (or increase a key size of a symmetric one) while ensuring that the existing data can be migrated to a new format. However, it may require downtime to re-encrypt existing data with a quantum-safe algorithm. 

Often, altering existing (legacy) systems to address the security concern may be challenging. Legacy systems frequently lack adequate information or support to be maintained or upgraded. The root causes of these issues are numerous. For example, the system developers may be unavailable (e.g., because they left the company or retired), source code or documentation may be lost, or build platforms for the source code may be sunset. To make matters worse, the encryption-related code may be spread or cloned among multiple software components (due to bad design), making alterations even more challenging. These root causes make it extremely difficult and expensive to upgrade such a system to the newest security protocols, sometimes making the replacement the only feasible option. 

\subsection{Threats to the existing data}\label{sec:data}
As discussed in Section~\ref{sec:cta}, sensitive encrypted data (e.g., protected by encrypted network packets of TLS or an encrypted disk) are vulnerable to the harvest-then-decrypt attack, meaning one can leverage QC power to break the asymmetric encryption used by TLS or brute-force access to the encrypted disk and recover the sensitive data. While not much can be done about the protocols involving asymmetric algorithms; for the symmetric ones, we can increase the length of the key right away to counter the brute-force attack~\citep{Muppidi2018}. We can also encrypt archived data (e.g., stored on backup devices) with a quantum-safe algorithm.

Another example is a blockchain platform using proof-of-work algorithms (e.g., based on SHA-256). The security of blockchain platforms relies on a digital signature, which is based on either Elliptic Curve Digital Signature Algorithm~\citep{JohnsonMV01} or RSA algorithms; both are vulnerable to QCs. Kiktenko et al.~\citep{KiktenkoPATYKLF17} proposed a quantum-secured blockchain framework that utilizes QKD techniques via an experimental fiber network (the cost of the network is not disclosed). A more efficient approach may be to switch to a proof-of-stake algorithm, which does not require solving computationally intensive mathematical puzzles, from a proof-of-work algorithm.

\section{Our method}\label{sec:method}

We proposed a \textbf{7E} roadmap for software developers in~\citep{zhang2021quantum}, summarizing steps to address encryption-related challenges associated with quantum advantage and giving developers a structured way to prepare for the quantum advantage era. The 7E roadmap is inspired by the lessons learned from the Y2K era and is amalgamated with modern knowledge. We recap key features of the seven steps of 7E for completeness below. More detailed steps can be found in Section~\ref{sec:new7e}.

\begin{enumerate}
    \item \textbf{E}ngage executives and senior management so that they can sponsor the initiative. 

    \item \textbf{E}xamine existing products and their cybersecurity components to identify and locate the issues, review the document and programs and assess the problems. 

    \item \textbf{E}volve: design a new software with crypto-agility (as per NIST recommendations~\citep{chen2016report}) in mind so that quantum-safe algorithms can be added to the software later on. 
    
    \item \textbf{E}ducate the programmers and designers to ensure that everyone is ``on the same page'' because (in most cases) the security-related component is coupled with the remaining software components. 
    \item \textbf{E}stimate the impact of potential problems and the cost of alternatives to prioritize the problems. 

    \item \textbf{E}xecute the new cybersecurity policy. 

    \item \textbf{E}ssay the new cybersecurity policy. Keep monitoring the performance and the robustness of your new cybersecurity policy in production to make sure that the challenges associated with quantum advantage were addressed; adjust the policy if needed.
\end{enumerate}

We keep the generality of the 7E roadmap so that it serves as a general guideline and can be generalized in other applications. However, while the 7E roadmap shares some common characteristics with other software evolution guidelines (especially, ones in regard to Y2K), we highlight two distinguishing characteristics of 7E roadmap for quantum-safe cryptography:
\begin{enumerate}
    \item  \textbf{Crypto-agility}, which is important for PQC evolution because NIST is still standardizing post-quantum cryptography~\citep{nist_timeline}, and a FTQC has not yet been developed. Thus, the current implementations of PQC may be changed in the future. Crypto-agility provides the ability to evolve the cybersecurity components without significant impact on the rest components of your system;
    \item \textbf{Education}, as the challenges of quantum attacks are new to managements and developers. Thus, knowledge transfer is important to make sure everyone is on the same page. 
\end{enumerate}
Additional details about how 7E differs from generic guidelines for maintaining and evolving a software product are provided in Section~\ref{sec:lessons}.

\section{Enhanced 7E Roadmap}\label{sec:new7e}

To validate the 7E roadmap, we conduct a case study on IBM Db2 database system (hereon abbreviated to Db2). Based on our case study (details of which will be given in Section~\ref{sec:industry}-onward), we incorporate our experience\footnote{Please note that not all of our experience will necessarily apply to other cases. However, by sharing as much information as possible, we allow practitioners and researchers to decide which approaches are best for them.} and findings into the original 7E roadmap and propose an enhanced 7E roadmap. The key enhancements are highlighted in \textbf{bold} text. In addition, we find out the importance of education. Thus, we bring Step 4 of ``Educate'' to Step 2 in the enhanced version.

\begin{enumerate}
    \item \textbf{E}ngage executives and senior management so that they can sponsor the initiative. It is important to get acknowledgement from the decision makers in your company or organization. Moreover, executives and senior management can assess security concerns from a broader perspective. They will also sponsor the allocation of the human resources (e.g., the experts within the organization who will drive the project forward).  Quantum computing is an emerging technology, and it may take time for executives to learn its growing powers and realize the side effect of these powers on the existing cryptography schemes. To educate the management, you can use formal presentations or reports and incorporate their feedback later. 
    
    \textbf{In addition to management, we also need to engage multiple experts from different domains\footnote{In our case, we have a CTO of Data Security (IBM), Senior Software Developer (IBM Db2), Architects (IBM Security), and Software Engineering Researchers (Ryerson University).}, because the evolution of the cybersecurity component often involves other parts of the system. For our project to be successful, multiple experts had to collaborate.}

    \item \textbf{E}ducate the programmers and designers to ensure that everyone is ``on the same page'' because (in most cases) the security-related component is coupled with the remaining software components. This implies that the whole development organization needs to be aware of the challenges of quantum attacks. The application of PQC is still in its infancy. Thus, security professionals may not have sufficient knowledge of PQC to perform the evolution of cryptographic schemes confidently. This step is complementary to Step~1, focusing on training technical staff rather than the management. 
    
    \textbf{We transfer knowledge and brainstorm within and outside this research team through work sessions, conferences, seminars, publications, etc. Researchers and practitioners can use these activities to estimate the impact of PQC evolution on IBM Db2 software. In addition, discussions during these events generate new ideas and questions\footnote{For example, choosing the right version of Kyber, so that performance and security are balanced.}.}

    \item \textbf{E}xamine existing products and their cybersecurity components to identify and locate the issues, review the document and programs and assess the problems. For legacy systems, there exist difficult scenarios, such as lack of documentation, source code, or build infrastructure (as discussed in Section~\ref{sec:impact}). Identify existing data (if any) that may require protection. 
    
    \textbf{Note that the evolution of PQC may also affect other components of the existing system. As an example, our work mainly altered the user authentication component of IBM Db2. Nevertheless, there are some cases where we need to make changes outside of the user authentication component. For example, some checks of encryption key sizes have been found in other parts of Db2.}

    \item \textbf{E}volve: design a new software with crypto-agility (as per NIST recommendations~\citep{chen2016report}) in mind so that quantum-safe algorithms can be added to the software later on. Conceptually, this should be achieved by designing the systems in such a way that an existing encryption scheme can be easily replaced with a new one. Moreover, the systems should be able to recognize and ``translate'' multiple encryption schemes.
    
     For example, the encryption component can be designed to be plug-and-play so that an existing encryption algorithm can be replaced (if it is discovered to be vulnerable) with a robust one. This will save costs in the future when the standards of PQC are finalized, and our software has to be updated with these algorithms. 
     \textbf{We achieve crypto-agility to a certain degree by replacing fixed key size with dynamic key size in our case study. At the same time, we acknowledge the challenges to redesign the whole encryption component of existing software to implement crypto-agility.}
    
    The challenges (identified in the previous steps and Section~\ref{sec:impact}) may affect the easiness of design decisions. For example, if documentation is missing, one may first have to reverse engineer the architecture from the code by extracting caller-callee function pairs and identifying security-related ones.
    
    For cryptography that involves multiple business partners, achieving crypto-agility requires that all business partners update hardware and software promptly. Moreover, all the partners should disclose crypto-related information, e.g., security certificates and protocols, to each other. Not only that, multi-organizational interactions can be enabled by institutionalizing open standards, preferably with open source reference implementations of those standards. Conceptually, this process would be akin to the creation of other open standards, where an open standards organization would be formed to drive the process. 

    \item \textbf{E}stimate the impact of potential problems and the cost of alternatives to prioritize the problems. The findings from Steps 2 and 3 should help to estimate the cost. Rate the cost of potential solutions in terms of human and time resources. Work first on the systems that handle critical data first. The definition of critical will vary from industry to industry. A representative example is a system handling personal data, such as financial transactions or health records. \textbf{It is always a good idea to prepare some buffer time in your project management. Though our project involved multiple experts and had a clear roadmap, we still underestimated the time needed to complete it by about $50\%$ due to the challenges discussed in Section~\ref{sec:lessons}.}
    
    Upgrade of the cryptographic schemes involving symmetric encryption will typically be cheaper than the asymmetric one. The former will usually require increasing the length of the key and increasing the storage space for the keys. The latter may also require replacing the code needed to do the encryption.

    \item \textbf{E}xecute the new cybersecurity policy. Select and adopt appropriate solutions based on requirements, budgets, and priorities. As discussed in Section~\ref{sec:impact}, practitioners can execute the new cybersecurity policy in different ways. For newly-built systems, PQC may be adopted (see Sections~\ref{sec:asymmetric} and~\ref{sec:symmetric}). For legacy systems, the software and associated hardware may have to be altered (see Section~\ref{sec:impact}). For existing data, an intermediate solution~--- e.g., re-encrypting the existing data with a quantum-safe cryptographic algorithm~--- may be applied (see Section~\ref{sec:data}). \textbf{We choose the latter because we work on the existing Db2 software. The challenges and take-away messages from our execution can be found in Section~\ref{sec:lessons}.}

    \item \textbf{E}ssay the new cybersecurity policy. Keep monitoring the performance and the robustness of your new cybersecurity policy in production to make sure that the challenges associated with quantum advantage were addressed; adjust the policy if needed. Experiences and lessons learned from one project may also apply to another one. These lessons could serve as a building block to a general theory of making PQC evolution agile and smooth. \textbf{We transfer our knowledge of PQC and the experience with Db2 both inside and outside IBM via inter-project discussions, conferences, seminars, publications, etc. We hope our roadmap and experience can help more practitioners get prepared for the year of quantum advantage (also known as Q2K).}
\end{enumerate}

\section{Industrial Settings}\label{sec:industry}

We choose Db2 because it complies with various security standards to protect sensitive data. In addition, it is a mature product (initially released in 1987, 34 years ago). However, it is under active development, with new features being added regularly. Db2 codebase consists of tens of millions of lines of C/C++ code (note that the C/C++ codebase was started in the early 1990s). 
Let us briefly discuss Db2 features that require cryptographic algorithms (that are vulnerable to quantum attacks) and identify how to replace them with PQC algorithms.

Db2 consists of a Db2 server and data server clients. A Db2 server is an enterprise-grade relational database management system that delivers data to clients. A Db2 client is an application that runs Db2 and SQL commands against the server.

The first step of accessing a Db2 database is authentication. Authentication is the process by which a system verifies a user's identity. According to IBM Db2 documentation version 11.5~\citep{Db2115IB0:online}, Db2 requires two items to authenticate a user, i.e., a user ID and a password. The user's identity is verified through the correct user ID and its corresponding password. 

\begin{figure}[!th]
	\centering
	\includegraphics[width=0.7\columnwidth]{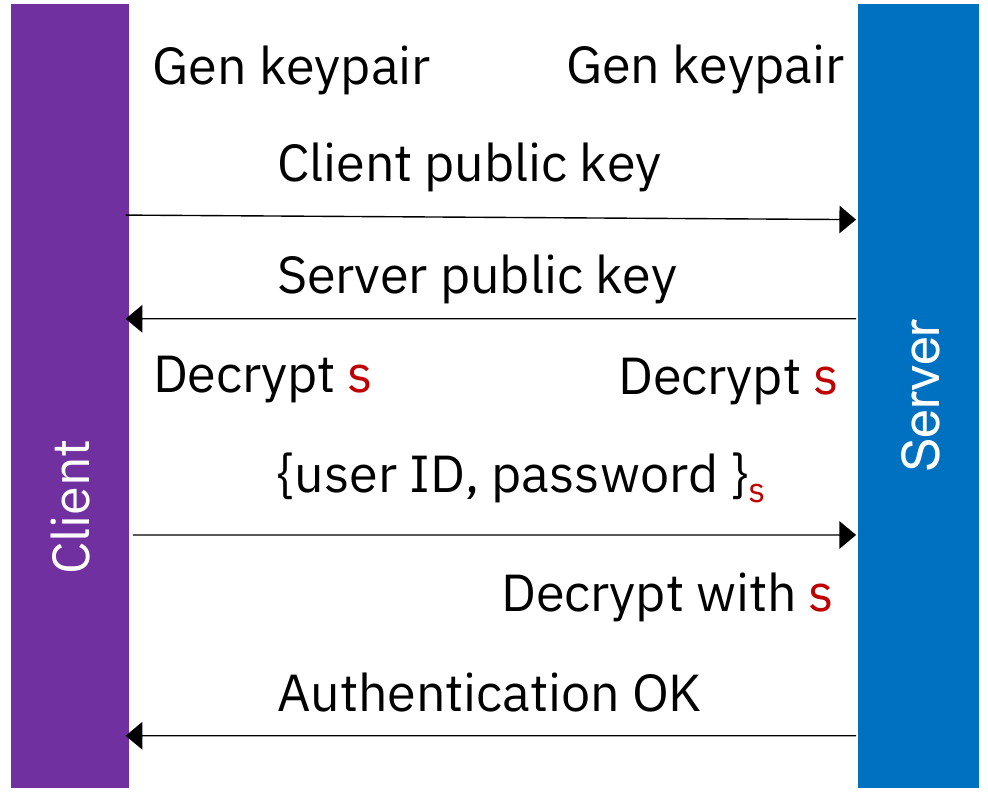}
	\caption{DH key exchange method establishes a shared secret between two parties for authentication. Both the client and the server generate a pair of keys, i.e., private key and public key. They then exchange their public keys. The client and the sever independently compute the shared secret (\textit{s}) using the other's public keys and their own private keys. The client encrypts user ID and password using \textit{s} and sends them to the server. After decrypting the user ID and password using \textit{s}, the server finishes the authentication of the client.}
	\label{fig:dh}
\end{figure}

To keep in-transit data safe, Db2 builds an encrypted communication between the user and Db2 using TLS protocol. TLS uses both asymmetric cryptography (e.g., DH) and symmetric cryptography (e.g., AES) for encryption.\footnote{Technical note: the implementations of DH and TLS are independent. DH is adopted at the Distributed Relational Database Architecture (DRDA) layer~\citep{Distribu58:online}, i.e., the application layer (layer seven) in Open Systems Interconnection (OSI) model~\citep{hebrawi1993open}. TLS is implemented at the transport layer (layer four) in OSI and is independent of the DRDA layer.} DH is a method to exchange cryptographic keys over a public channel securely. Db2 adopts DH to exchange a shared secret key between the user and Db2. The key is then used to encrypt subsequent communications with symmetric cryptography. Figure~\ref{fig:dh} illustrates a simplified process of DH key agreement between two parties for authentication. 

\textbf{Key exchange algorithm.} DH key exchange relies on the assumption that discrete logarithm problem is hard to solve on classical computers (as discussed in Section~\ref{sec:asymmetric}). However, the discrete logarithm problem can be efficiently solved using Shor’s algorithm on quantum computers. In other words, DH key exchange is vulnerable to quantum attacks. To tackle this problem, we need to find a quantum-safe replacement.

\begin{figure}[!th]
	\centering
	\includegraphics[width=0.7\columnwidth]{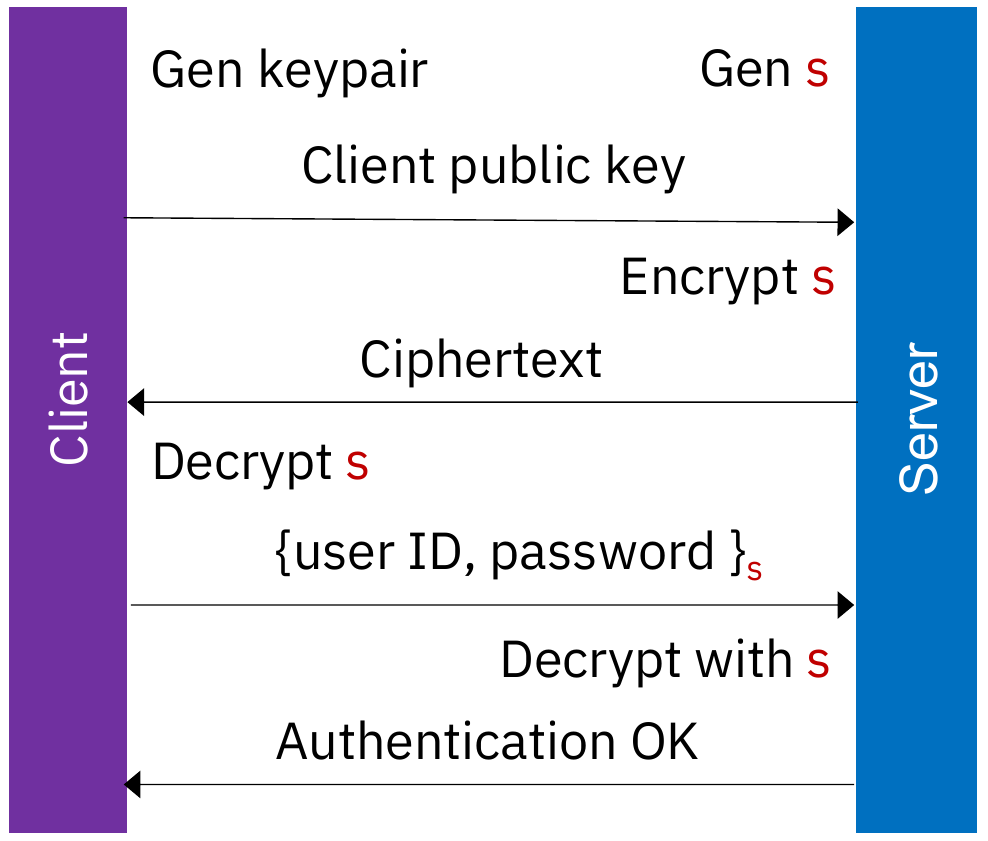}
	\caption{Kyber key encapsulation mechanism between two parties for authentication. The client generates a pair of private key and public key, and the server generates a random number as the shared secret (\textit{s}). The client sends its public key to the server. The server uses the client's public key to encrypt \textit{s} into a ciphertext and sends the ciphertext back to the client. The client derives \textit{s} from the ciphertext using its own private key. Then, the client encrypts its user ID and password using \textit{s} and sends them to the server. After decrypting the user ID and password using \textit{s}, the server finishes the authentication of the client.}
	\label{fig:kyber}
\end{figure}

We choose Kyber~\citep{bos2018crystals}, which is one of the algorithms in the Cryptographic Suite for Algebraic Lattices (CRYSTALS), as the successor of DH. Kyber is a key encapsulation algorithm based on the hardness of solving the learning-with-errors problem over module lattices~\citep{Kyber60:online}. Kyber is one of the four public-key encryption and key-establishment algorithms in the round three finalists announced by NIST~\citep{PostQuan59:online}. The procedure of Kyber key exchange in user authentication can be seen in Fig.~\ref{fig:kyber}.

\textbf{Digital signature and TLS.} Since DH key exchange does not provide authentication, it is vulnerable to man-in-the-middle attacks. For this reason, it is implemented alongside RSA digital signature for authentication in TLS. However, as discussed in Section~\ref{sec:intro}, RSA is also vulnerable to quantum attacks. Thus, we need to replace RSA with a quantum-safe digital signature scheme.

We choose Dilithium~\citep{Dilithiu84:online}, also part of CRYSTALS, as our new digital signature scheme. Similar to Kyber, Dilithium is also based on the hardness of lattice problems over module lattices. Dilithium is one of the three finalists of digital signature algorithms submitted to NIST in round three~\citep{PostQuan59:online}. 

TLS~1.3 offers significant improvements over its predecessor TLS~1.2; the key feature that interests us is its support of post-quantum authentication within IBM Global Security Kit (GSKit)~\citep{IBMGloba90:online}. For this reason, we upgrade\footnote{To do the upgrade, we integrated Dilithium into TLS~1.3.} TLS~1.2 in Db2 to TLS~1.3.

\section{Experiments}\label{sec:experiment}

In this section, we discuss the implementation of Kyber, Dilithium, and TLS~1.3 in Db2.

\subsection{GSKit}

IBM GSKit provides libraries and utilities for both general purpose cryptography and Secure Sockets Layer (SSL) or TLS communication. More specifically, GSKit consists of two separate packages, containing different functionalities~\citep{Introduc20:online}
\begin{enumerate*}
    \item GSKit Crypt, which contains the cryptographic algorithms that the package of GSKit SSL depends on; 
    \item GSKit SSL, which includes the essential runtime support to enable security calls and the use of the TLS protocol.
\end{enumerate*}

In our experiments, we focus on two things 
\begin{enumerate*}
    \item use of the general-purpose cryptographic library to evolve the algorithm used in Db2 user authentication towards quantum safety (i.e., we upgrade from DH to Kyber), and
    \item use of the TLS implementation to evolve the Db2 TLS capability towards quantum safety (i.e., we migrate from TLS~1.2 to TLS~1.3).
\end{enumerate*}

\subsection{Implementation of Kyber}

As can be seen in Fig.~\ref{fig:kyber}, Kyber key encapsulation scheme shares a secret key between the server and the client in three key steps as follows.

\begin{enumerate}
    \item Client generates a private/public Kyber key pair and sends the public key to the server.
    \item Server takes a public key and uses a random seed to generate a shared key and ciphertext.
    \item Client uses private key and ciphertext to get the shared key.
\end{enumerate}

After these three key steps, we can use Kyber shared secret as a symmetric key (e.g., an AES 256 bit key) to protect channel between client and server.

\lstset{language=C}
\begin{lstlisting}[caption={Kyber APIs in GSKit.},label={lst:gskit},captionpos=b,frame=single,float=tp,floatplacement=tbp,basicstyle=\footnotesize]
//Keypair generation API (client)
int crypto_kem_keypair (unsigned char *pk, 
                        unsigned char *sk, 
                        KYBK k);

//Key encryption API (server)
int crypto_kem_enc (unsigned char *ct, 
                    unsigned char *ss, 
                    const unsigned char *pk, 
                    KYBK k);

//Key decryption API (client)
int crypto_kem_dec (unsigned char *ss, 
                    const unsigned char *ct, 
                    const unsigned char *sk, 
                    KYBK k);
\end{lstlisting}

GSKit implements Kyber and provides three APIs corresponding to the three key steps above. They can be found in Listing~\ref{lst:gskit}.

The parameter \code{pk} is public key, \code{sk} is secret key, \code{ct} is ciphertext, \code{ss} is shared secret, and \code{k} is Kyber key, which indicates the security level of Kyber~\citep{Kyber60:online}.  

\subsection{Dilithium and TLS~1.3}

TLS, which is the successor of SSL, provides communications security over a computer network. Statistics show that over 90\% of the web traffic is based on SSL/TLS-based HTTPS protocol~\citep{HTTPSenc30:online}. Db2 uses TLS to secure all communications between servers and clients. TLS protocol requires digital signatures (in most cases, based on asymmetric cryptography). 

In TLS, symmetric cryptography, i.e., Data Encryption Standard (DES) or AES, is used to encrypt the data transmitted. The keys for this symmetric encryption are uniquely generated for each connection and are based on a shared secret, which was negotiated using a public-key-agreement protocol --- DH. The identity of the communicating parities are authenticated using asymmetric cryptography, i.e., RSA.

As discussed in Section~\ref{sec:industry}, we need to replace RSA with quantum-safe cryptography, i.e., Dilithium. To enable a post-quantum digital signature scheme, we need to upgrade TLS to the latest version --- 1.3 --- from the existing version --- 1.2. GSKit has implemented PQC algorithms as an extension of TLS 1.3 protocol. 

The upgrade of TLS consists of two steps. In the first step, we need to change Db2 codebase so that Db2 will use TLS~1.3 provided by GSKit. This change is associated with two settings: 
\begin{enumerate*}
    \item enable TLS~1.3 with PQC (on both client and server sides) for the PQC handshake to take place,
    \item enable PQC certificates in TLS~1.3 for a digital signature on the server side. 
\end{enumerate*}
In the second step, we need to update the certificate chain using the GSKit command tool \code{gsk8capicmd} for key and certificate management. We create a traditional certificate chain and an attached post-quantum certificate chain to create the post-quantum certificates for TLS~1.3. 

To verify the implementation of Dilithium and TLS~1.3, we print out the HTTPS connection reports and the TLS handshake transcript. The HTTPS connection reports verify two things:
\begin{enumerate*}
    \item TLS~1.3 has been enabled, and 
    \item quantum-safe certificates are used. 
\end{enumerate*}
To confirm the result, the TLS handshake transcript is used to verify that PQC, i.e., Dilithium, has been enabled (instead of RSA). 

\subsection{Performance evaluation}

\begin{figure}[tb]
    \centering
    \includegraphics[width=\columnwidth]{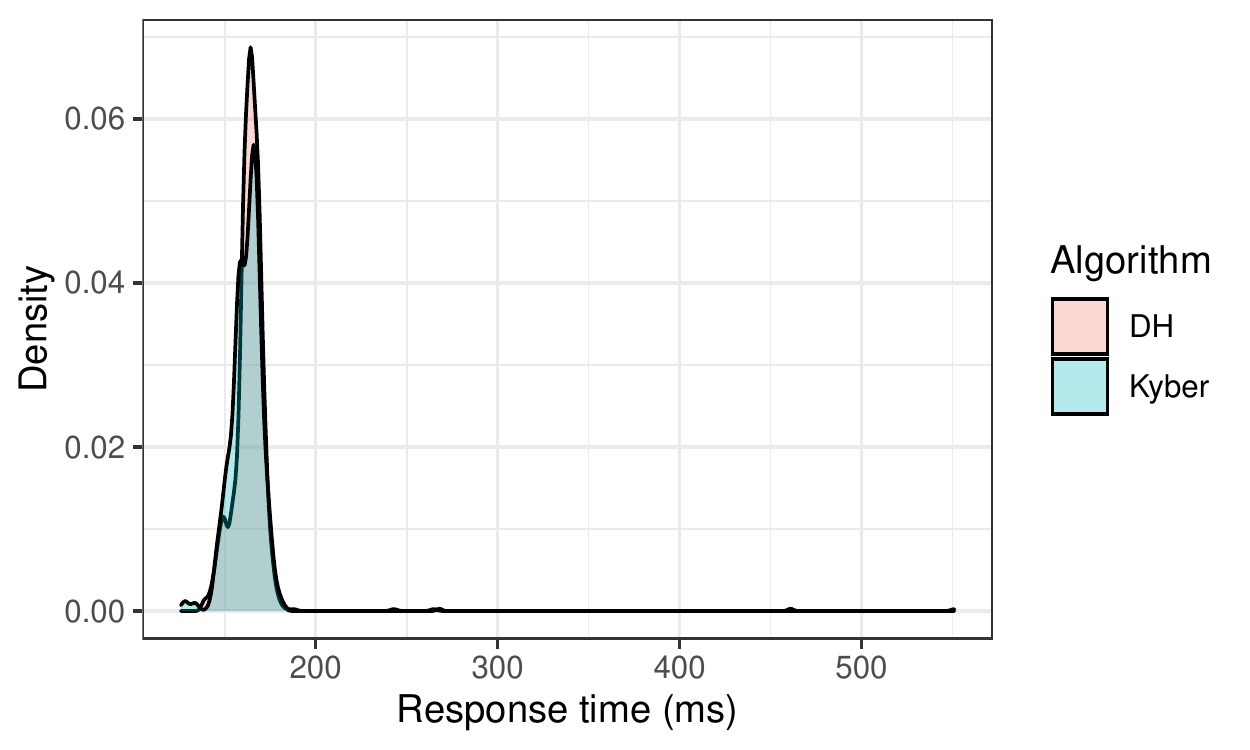}
    \caption{Distribution of response time.}
    \label{fig:timing}
\end{figure}

We evaluate the performance difference between Kyber and DH on our test bed, which has 32 CPUs (1.2 GHz) and 256 GB of memory. The performance is evaluated by running the command of \code{db2 connect} from a client to connect to a database on the server, which triggers TLS communication and user authentication. To evaluate the performance of Kyber, we adopt the Monte Carlo method to repeat this test 1,000 times with both Kyber and DH, respectively. The distributions of the timing data are shown in Fig.~\ref{fig:timing} and summary stats --- in Table~\ref{tab:perf}. The average response time with Kyber is 162.514 ms, and the standard deviation is 15.281 ms; the average response time with DH is 163.552 ms, and the standard deviation is 12.222 ms. Based on the QQ-plots and Shapiro-Wilk normality test, the distributions of timings are non-normal. The difference between the distributions are statistically significant: the one- and two-sided Wilcoxon rank-sum tests yield $p$-value of $\approx 0.0002$ and  $\approx 0.0004$, respectively. From a practical perspective, we can conclude that the performance of Db2 is not affected by the PQC upgrade --- the average response time even decreases by 0.635\% after we migrate from DH to Kyber (which is vital to performance-critical systems).

\begin{table}
\caption{Summary stats of response time.}
\label{tab:perf}
\centering
\begin{tabular}{l|r|r}
\toprule
Algorithm & Avg. response time (ms) & St. Dev. (ms) \\ \midrule
Kyber & 162.514  & 15.281  \\
DH & 163.552  & 12.222  \\ \bottomrule
\end{tabular}
\end{table}

\section{Challenges and take-away messages}\label{sec:lessons}

We list the challenges that we met during the development and the lessons learned from this project below.

\begin{enumerate}
    \item 
    Lack of documentation, ambiguity of documentation, and obsolete documentation (corresponding to Step 2 in 7E --- Exam). Because of the passing of time, mature systems often lack development documentation, or the documentation is not up-to-date, which slows down the pace of the learning curve for the cryptography system. Even if the document exists, it can contain errors. \\ 
    For example, we uncovered errors in the API document, where the socket handle parameter should have been an environment handle. \\ 
    Another example relates to outdated documents. In one case, the instructions in the document we received were written for 32-bit operating systems, but we were running on the 64-bit system. As a result, the commands provided in the document were inoperable, and we had to explore and create new instructions tailored for the modern operating system.

    \item 
    Legacy/fragile development environment (corresponding to Step 6 in 7E --- Execute). Because of the long development history of Db2 (34 years) and its mission-critical feature, Db2 project management requirements often lead to a limited number of unified legacy solutions. Many of them are command-line tools. Compared to graphical user interface, complex command-line interface has a steeper learning curve. 

    \item 
    Large codebase (corresponding to Step 2 in 7E --- Exam). A large codebase (in our case, tens of millions of lines of C/C++code) with a long history of development often has three drawbacks during maintenance or evolution, i.e., 
    \begin{enumerate*}
        \item lack of comments,
        \item complex structures, and
        \item long compilation time.
    \end{enumerate*}
     Let us elaborate on these three items now. 
     First, the lack of well-written comments (detailing the whole design) can make code evolution much more challenging. Second, as the size of the codebase increases, the project becomes more and more complex. Consequently, the project becomes less readable and understandable, making it more complicated to apply new changes to the system~\citep{jones2008applied}. Finally, because of the codebase size, compilation can take a long time to finish. For instance, a complete build (i.e., recompiling the whole codebase from scratch) for Db2 takes about two to three hours on our test bed. This is an architectural necessity in Db2 because the crypto code resides in the components that many other higher-level components depend on.

    \item 
    Distributed teams (corresponding to Step 1 in 7E --- Engage). Upgrading a complex system often involves collaborations among multiple teams or organizations. In this project particularly, we require expertise in two different areas, i.e., database engine (expert in the security component) and cryptography (research team and development team). Since the component that we need to upgrade in Db2 is mainly security related (i.e., the implementations of DH and TLS in Db2), we require expertise in Db2 security. 
    During the PQC implementation, communication and collaboration among multiple geo-distributed teams (with different backgrounds) are required. A breakdown or lag in communication could cause delays in the development. 

    \item 
    Technical debt because of hard coding (corresponding to Step 3 in 7E --- Evolve). Technical debt is a concept in software engineering that reflects the extra work caused by previous work when choosing an easy solution instead of applying the best overall solution~\citep{WhatisTe38:online}. Hard coding is one of the most common decisions that lead to technical debt. 
    Let us consider the case that we met in the evolution of encryption schemes --- hard-coded cryptographic key lengths. Whenever you are required to increase the encryption key length (e.g., from 128-bit key to 256-bit key), to increase the level of cryptographic complexity, you are exposed to the risk of exceeding the bound of the key length (usually implemented in an array), leading to memory corruption and system crash.\footnote{In C programming language, the compiler does not automatically check array bounds. Your array index may point to a virtual memory that is not mapped to a physical memory, which can cause a crash.} 
    In our project, not only the key lengths are hard-coded, but there also exist multiple and inconsistent key length checks, e.g., an ``if'' condition where key length has to be smaller than a specific constraint. In such a case, we need to loosen the constraints. However, those key length constraints are hard to trace because the key names can be changed, and the constraints are not necessarily at the same security layer (e.g., cryptographic protocol level vs. algorithm level).

    \item 
    Legacy encryption scheme on the client-side (corresponding to Step 3 in 7E --- Evolve). As mentioned before, Db2 has been evolving for 34 years. Due to its long history, Db2 may still need to support older database client drivers that some customers may still be using in legacy environments. For example, the database security administrator may configure the Db2 server to allow DES instead of AES, typically in a tightly controlled environment. 
    For various reasons (e.g., compatibility and/or cost issues), those customers keep using legacy software. To support those customers, not only do we need to maintain the compatibility of the security component for old versions of the software, but we also need to support customers when transitioning to new versions of the software. 
    
    \item 
    Cryptographic APIs are in flux (corresponding to Step 6 in 7E --- Execute). 
    Both Kyber and Dilithium are still in development (NIST has not finalized the PQC standardization). To the best of our knowledge, the only commercial product that adopts Kyber (based on the version submitted to NIST competition Round 2) is Amazon Web Services Key Management Service~\citep{Round2po31:online}. At IBM, the applications of Kyber and Dilithium are not yet commercially supported. The Application Programming Interfaces (APIs) are not released for production teams. Moreover, for security purposes, the API functions to access the PQC functionality are deliberately hidden by the developers of the cryptographic libraries (as they want to make sure that beta-level code does not accidentally end up in the production code), which makes it difficult for developers of the main product to integrate the functionality. 
    In our case, we required technical support from the cryptography research team and GSKit development team to apply the APIs of Kyber and Dilithium to Db2. We first spent our time with the developers of the cryptographic libraries in online meetings trying to understand how to access the API functions. Then, we wrote the wrappers implementing the tricks needed to access the hidden functions. Once the cryptographic library APIs become officially available, the developers of the main product will have to replace the customized wrappers with the direct calls to the functions (which will not be a risky procedure, as long as the API calls do not change drastically).

    \item 
    Underestimation of sizing even though correct experts are brought in (corresponding to Step 5 in 7E --- Estimate). As a cutting-edge and ever-changing technology, the application of PQC is still in its infancy, and developers have a steep learning curve for the application of PQC. In our project, though we involved multiple top experts at IBM, we realized that we had underestimated the size of the project (in terms of work time) during the development. 
    While we hope that readers will be more accurate in their sizing estimates, we know that project size underestimation is a common issue in Software Engineering. We recommend that the readers err on the side of caution when estimating the amount of effort and resources required for PQC evolution.
    
    \item 
    An iterative and incremental approach is preferred to the waterfall-like model (corresponding to Step 2 in 7E --- Exam). During the development, we realized that the uncertainties associated with our new code changes often require more changes in other components of the system. For instance, a small component change may lead to significant impact on the performance and/or functionality of the whole system. The reasons behind this are numerous: e.g., lack of understanding, lack of documentation, or code complexity. Thus, we advise adopting iterative and incremental methods in PQC evolution. In our case, we separate the PQC-related code into several modules and add these modules incrementally. We test the cryptographic functions and performance incrementally during the development.
    
    \item 
    Education is essential (corresponding to Step 4 in 7E --- Educate). The education went both ways. The research part of the team educated the technical part of the team about the PQC-related challenges. In contrast, the technical experts educated researchers on the issues specific to the upgrade of the Db2 source code.
    
    \item 
     Maintain the PQC infrastructure (corresponding to Step 7 in 7E --- Essay). Once the PQC is in place, the next step is to maintain the PQC infrastructure and prevent developers from introducing new security vulnerabilities. To achieve this goal, one can implement a cryptographic algorithm/code scanning policy to monitor future changes of cryptographic algorithms and ensure no quantum-unsafe algorithms are accidentally added to the code base. Human and technical resources should be allocated to the task by the management.
    
\end{enumerate}

\textbf{Take-away messages.} Besides crypto-agility, we also mentioned other lessons learned in the previous section. From this experience, we summarize some takeaway messages. We would like to share these messages with practitioners who want to upgrade the current cryptographic algorithms to quantum-safe algorithms in their existing systems. We hope they will benefit from our suggestions. Our messages are as follows.

\begin{enumerate}
    \item Prepare ahead and have a clear roadmap and timeline. As  mentioned in Section~\ref{sec:method} and Section~\ref{sec:lessons} (Step 8), follow the 7E steps to update your existing cryptography and prepare for additional time for the development.
    
    \item Get support from management. As mentioned in Section~\ref{sec:method} (Step 1), PQC is a new emerging technology, it is important to educate your management and colleagues so that they are aware of the potential risk of data breach because of quantum attacks. It is more important to receive support from management so that the upgrade of the cryptography can be successfully proceeded with the collaboration from all the departments that need to be involved.
    
    \item Collaborate with multiple departments. As mentioned in Section~\ref{sec:lessons} (Step 4), the upgrade of cryptography can affect other components of the system or even cause failure of the system. Thus, we suggest that all stakeholders should be engaged in the upgrade as soon as possible to lower any risk to the system. Moreover, those experts will help to achieve the success of the project.
    
    \item Prepare to pay the hard-coding debt. As mentioned in Section~\ref{sec:lessons} (Step 5), technical debt can be a block in the way of your project. A thorough exam of the existing code is necessary to clean up the blocks. This is also one of the potential reasons that can cause delays in your project.
    
    \item Document the development for future maintenance and evolution. As mentioned in Section~\ref{sec:method} (Step 7) and Section~\ref{sec:lessons} (Step 10), NIST is still working on the standardization of PQC. This implies that PQC will evolve. Thus, it is a good idea to keep all the records of your development to simplify future changes.
    
    \item Plan for crypto-agility. The reason behind this is the same as the previous one --- the evolution of PQC. As mentioned in Section~\ref{sec:method} (Step 3) and Section~\ref{sec:lessons} (Step 5), we need to design the new cryptography with crypto-agility (such as dynamic key sizes). Paying off technical debt at this stage will improve the productivity in future development.
    
    \item Measure performance impact. As mentioned in Section~\ref{sec:method} (Step 7), we need to assess  the new cryptographic system's performance and robustness. For performance-centric software systems, our finding shows that a PQC upgrade does not increase the time required to exchange cryptographic keys (and, on average, may reduce the timing slightly). 
\end{enumerate}

\section{Discussions}\label{sec:discussions}

\textbf{Threats to validity.} We cannot generalize our findings to all software products~\citep{wieringa2015six}. However, Db2 can be treated as a ``critical case'' (in a case study sense of the term~\citep{yin2009case}), showing the potential viability of 7E and highlighting some of the potential pitfalls. Db2 uses widespread encryption schemes, thus, our experience may be useful to other developers of software products. The 7E roadmap can be readily applied to other projects; thus, we encourage researchers and practitioners to apply 7E and our lessons learned to other instances of PQC evolution and are looking forward to their findings.

In the last 12 months, we have witnessed increased attention to the risk posed by quantum computing. Some leading institutions have published guidance to help organizations transition to quantum-safe encryption. While our 7E guidance predates these recent publications, it is worth noting that in essence our 7E and this recently published guidance are very much in synchronization. For instance, IBM proposes a framework that consists of four pillars --- preparation, discovery, transformation, and observability~\citep{Security57:online}. Preparation maps to our ``Engage'' and ``Educate'' principles, discovery maps to our ``Examine'' principle, transformation maps to our ``Evolve'', ``Estimate'', and ``Execute'' principles, and lastly, observability maps to our ``Essay'' principle. Similarly, the US Cybersecurity \& Infrastructure Security Agency shares similar guidance that can also be grouped into three pillars --- preparation, discovery, and transformation~\citep{Preparef12:online}. While the naming may be different, the essence of our 7E is very much in synchronization with the guidance recently emerging. Consequently, these guiding documents may serve as indirect evidence of the generalizability and applicability of our findings.

\textbf{Further discussions.} As we expected, the introduction of the encryption schema is not a trivial process. Some of the pain-points are anticipated and are typical for the evolution of a large-scale legacy product. They are associated with difficulties in comprehending a massive codebase, as well as fragile source code management and build environments. 

However, some of the other problems are unexpected and are not typical. One example is the need to implement customized wrappers to call the PQC API functions because they are still in flux (see Challenge 7 in Section~\ref{sec:lessons}). Another example is the essentiality of education for people who are not aware of the urgency to ensure data security against quantum attacks (see Challenge 10 in Section~\ref{sec:lessons}). 

It is not easy to follow the crypto-agility paradigm strictly in existing legacy systems. In the previous section, we have shown that we improved the crypto-agility of Db2 by introducing flexible key sizes. However, we have not solved this problem completely. For example, the authentication code in Db2 is written in the C programming language for performance reasons. C language lacks object-oriented (OO) concepts, such as abstraction. As a result, it is difficult to implement more complex objects on top of an abstraction. Though one can introduce OO principles by rewriting the code to C++, it will be expensive, time-consuming, and may lead to performance degradation, violating essential performance requirements.

\section{Conclusions}\label{sec:conclusion}

This paper continued our work on the 7E roadmap for PQC evolution. While leveraging 7E, we conducted a case study to make IBM Db2 database management software quantum-safe. As a result, we validated the 7E roadmap throughout the project. Moreover, we pinpointed operational details needed to execute 7E, recorded lessons learned, delivered takeaway messages, and proposed an enhanced 7E for practitioners so that they could benefit from our experience and use it in their endeavours. Our data may also be of interest to academics as a building block in the theory of evolution of large mature software products.

For future work, we would like to apply the 7E roadmap and our best practices to other systems that are vulnerable to quantum attacks. At the same time, we would like to dig deeper into the topic of crypto-agility with PQC. One possible solution is to implement a hybrid classical-quantum cryptography. Finally, we would like to invite other practitioners and researchers to share their experiences with the usage of the 7E roadmap.

\bibliography{reference}

\end{document}